# Combined Theory of Complete Orthonormal Sets of Nonrelativistic, Quasirelativistic and Relativistic Sets of Wave Functions, and Slater Orbitals of Particles with Arbitrary Spin in Coordinate, Momentum and Four-Dimensional Spaces

I.I.Guseinov

*Department of Physics, Faculty of Arts and Sciences, Onsekiz Mart University, Çanakkale, Turkey*

**Abstract**

Using the complete orthonormal basis sets of nonrelativistic and quasirelativistic orbitals introduced by the author in previous papers for particles with arbitrary spin the new analytical relations for the $2(2s+1)$-component relativistic tensor wave functions and tensor Slater orbitals in coordinate, momentum and four-dimensional spaces are derived, where $s = 1/2, 1, 3/2, 2,\ldots$. The relativistic tensor function sets are expressed through the corresponding nonrelativistic and quasirelativistic orbitals. The analytical formulas for overlap integrals over relativistic tensor Slater orbitals with the same screening constants in coordinate space are also derived.

**Key words:** Relativistic and quasirelativistic tensor wave functions, Relativistic and quasirelativistic tensor Slater orbitals, Nonrelativistic scalar orbitals, Overlap integrals

## 1. Introduction

It is well known that the Schrödinger's nonrelativistic hydrogen-like orbitals and their extensions to momentum and four-dimensional spaces by Fock [1,2] are not complete unless the continuum is included. Hylleraas, Shull and Löwdin in Refs. [3-6] introduced in coordinate space the complete orthonormal sets of so-called Lambda and Coulomb Sturmian functions for the particles with spin s=0. Weniger [7] has shown that the Coulomb Sturmians form basis of a Sobolev space and, therefore, they are complete and orthonormal sets of functions. The method for constructing the relativistic one-electron Coulomb Sturmian basis set has been developed by Avery and Antonsen [8]. Vast majority of modern relativistic calculations are being done within numerically more convenient Gaussian type orbitals (GTO). In the case of the point nuclear model, as is well recognized in literature [9], GTO do not allow an adequate representation of important properties of the electronic wave function, such as the cusps at the nuclei [10] and the exponential decay at large distances [11]. Therefore, it is desirable to perform the relativistic calculations with the help of exponential type orbitals (ETO) when the point-like nuclear model is used. It should be noted that the straightforward use of relativistic equations in the calculations with finite basis sets, independent for large and small components, could cause numerical instability known as the Brown-Ravenhall disease [12]. A possible solution of this problem is to use "kinetically balanced" basis sets for the small components [13].

In Ref. [14] we have developed the method for constructing in coordinate, momentum and four-dimensional spaces the complete orthonormal sets for quasirelativistic tensor wave functions, and tensor Slater orbitals for particles with arbitrary spin s, where $s = 1/2, 1, 3/2, 2,...$, using corresponding scalar wave functions determined by the following relations:

for $\psi^\alpha$-exponantial type orbitals ($\psi^\alpha - ETO$)

$$\psi^\alpha_{nlm_l}(\zeta, \vec{r}) = R^\alpha_{nl}(\zeta, r) S_{lm_l}(\frac{\vec{r}}{r}) \tag{1}$$

$$\overline{\psi}^\alpha_{nlm_l}(\zeta, \vec{r}) = \overline{R}^\alpha_{nl}(\zeta, r) S_{lm_l}(\frac{\vec{r}}{r}), \tag{2}$$

for $\phi^\alpha$- momentum space orbitals ($\phi^\alpha - MSO$)

$$\phi^\alpha_{nlm_l}(\zeta, \vec{k}) = \Pi^\alpha_{nl}(\zeta, k) \tilde{S}_{lm_l}(\frac{\vec{k}}{k}) \tag{3}$$

$$\overline{\phi}^\alpha_{nlm_l}(\zeta, \vec{k}) = \overline{\Pi}^\alpha_{nl}(\zeta, k) \tilde{S}_{lm_l}(\frac{\vec{k}}{k}), \tag{4}$$

for $z^\alpha_{nlm_l}$- hyperspherical harmonics ($z^\alpha - HSH$)

$$z^\alpha_{nlm_l}(\zeta, \beta\theta\varphi) = P^\alpha_{nl}(\kappa_4) \tilde{T}_{lm_l}(\kappa_1, \kappa_2, \kappa_3) \tag{5}$$

$$\overline{z}^\alpha_{nlm_l}(\zeta, \beta\theta\varphi) = \overline{P}^\alpha_{nl}(\kappa_4) \tilde{T}_{lm_l}(\kappa_1, \kappa_2, \kappa_3), \tag{6}$$

for $\chi$- Slater type orbitals ($\chi - STO$)

$$\chi_{nlm_l}(\zeta, \vec{r}) = R_n(\zeta, r) S_{lm_l}(\frac{\vec{r}}{r}), \tag{7}$$

for Slater type u- momentum space orbitals ($u - MSO$)

$$u_{nlm_l}(\zeta, \vec{k}) = Q_{nl}(\zeta, k) \tilde{S}_{lm_l}(\frac{\vec{k}}{k}), \tag{8}$$

for Slater type $v$- hyperspherical harmonics ($v - HSH$)

$$v_{nlm_l}(\zeta, \beta\theta\varphi) = \frac{1}{\zeta^{3/2}} \Gamma_{nl}(\kappa_4) \tilde{T}_{lm_l}(\kappa_1, \kappa_2, \kappa_3). \tag{9}$$

See Refs. [15-18] for the exact definition of quantities occurring on the right-hand sides of Eqs. (1)-(9). We notice that the Cartesian coordinates $\kappa_1, \kappa_2, \kappa_3$ and $\kappa_4$ on the four-dimensional space occurring in Eqs. (5), (6) and (9) are obtained from the components of the momentum vector $\vec{k}$ by the following relations:

$$\begin{aligned} \kappa_1 &= k_x / \sqrt{\zeta^2 + k^2} = sin\beta cos\varphi sin\theta \\ \kappa_2 &= k_y / \sqrt{\zeta^2 + k^2} = sin\beta sin\varphi sin\theta \\ \kappa_3 &= k_z / \sqrt{\zeta^2 + k^2} = sin\beta cos\theta \\ \kappa_4 &= \zeta / \sqrt{\zeta^2 + k^2} = cos\beta, \end{aligned} \tag{10}$$

where the angles $\beta, \theta, \varphi$ ($0 \leq \beta \leq \frac{\pi}{2}$, $0 \leq \theta \leq \pi$, $0 \leq \varphi \leq 2\pi$) are spherical coordinates on the four-dimensional unit sphere; $\theta$ and $\varphi$ have the meaning of the usual spherical coordinates in momentum space. The surface element of the four-dimensional sphere is determined by

$$d\Omega(\zeta,\beta\theta\varphi) = \zeta^3 d\Omega. \quad (11)$$

This surface element is connected with the volume element in momentum space by the relation:

$$d^3\vec{k} = dk_x dk_y dk_z = d\Omega(\zeta,\beta\theta\varphi), \quad (12)$$

where

$$d\Omega = d\Omega(1,\beta\theta\varphi) = \frac{\sin^2\beta}{\cos^4\beta} d\beta \sin\theta d\theta d\varphi. \quad (13)$$

In this study, using functions (1)-(9) we obtain in coordinate, momentum and four-dimensional spaces a large number of relativistic tensor wave functions and tensor Slater orbitals. We notice that the relativistic tensor wave functions obtained are complete without the inclusion of the continuum.

## 2. Relativistic Tensor Wave Functions and Tensor Slater Orbitals

In order to derive the formulas for relativistic tensor wave functions and tensor Slater orbitals for particles with arbitrary spin in coordinate, momentum and four-dimensional spaces we use the corresponding quasirelativistic and nonrelativistic relations in the following form (see Ref. [14]):

For quasirelativistic tensor wave functions

$$K_{njm_j}^{\alpha ls} = \begin{pmatrix} a_{jm_j}^{ls}(0) k_{nlm_l(0)}^{\alpha} \\ a_{jm_j}^{ls}(1) k_{nlm_l(1)}^{\alpha} \\ \vdots \\ a_{jm_j}^{ls}(2s) k_{nlm_l(2s)}^{\alpha} \end{pmatrix}. \quad (14)$$

$$\bar{K}_{njm_j}^{\alpha ls} = \begin{pmatrix} a_{jm_j}^{ls}(0) \bar{k}_{nlm_l(0)}^{\alpha} \\ a_{jm_j}^{ls}(1) \bar{k}_{nlm_l(1)}^{\alpha} \\ \vdots \\ a_{jm_j}^{ls}(2s) \bar{k}_{nlm_l(2s)}^{\alpha} \end{pmatrix}, \quad (15)$$

where

$$K_{njm_j}^{\alpha ls} = \Psi_{njm_j}^{\alpha ls}(\zeta,\vec{r}), \Phi_{njm_j}^{\alpha ls}(\zeta,\vec{k}), Z_{njm_j}^{\alpha ls}(\zeta,\beta\theta\varphi) \quad (16)$$

$$\bar{K}_{njm_j}^{\alpha ls} = \bar{\Psi}_{njm_j}^{\alpha ls}(\zeta,\vec{r}), \bar{\Phi}_{njm_j}^{\alpha ls}(\zeta,\vec{k}), \bar{Z}_{njm_j}^{\alpha ls}(\zeta,\beta\theta\varphi). \quad (17)$$

For quasirelativistic tensor Slater orbitals

$$K_{njm_j}^{ls} = \begin{pmatrix} a_{jm_j}^{ls}(0) k_{nlm_l(0)} \\ a_{jm_j}^{ls}(1) k_{nlm_l(1)} \\ \vdots \\ a_{jm_j}^{ls}(2s) k_{nlm_l(2s)} \end{pmatrix}, \quad (18)$$

where

$$K_{njm_j}^{ls} = X_{njm_j}^{ls}(\zeta,\vec{r}), U_{njm_j}^{ls}(\zeta,\vec{k}), V_{njm_j}^{ls}(\zeta,\beta\theta\varphi) \quad (19)$$

and

$$n \geq 1, s \leq j \leq s+n-1, -j \leq m_j \leq j,$$

$$j - s \leq l \leq \min(j+s, n-1),$$

$$m_l(\lambda) = m_j - s + \lambda, \ 0 \leq \lambda \leq 2s.$$

For complete orthonormal sets of nonrelativistic scalar orbitals

$$k_{nlm_l}^{\alpha} = \psi_{nlm_l}^{\alpha}(\zeta,\vec{r}), \phi_{nlm_l}^{\alpha}(\zeta,\vec{k}), z_{nlm_l}^{\alpha}(\zeta,\beta\theta\varphi) \quad (20)$$

$$\bar{k}_{nlm_l}^{\alpha} = \bar{\psi}_{nlm_l}^{\alpha}(\zeta,\vec{r}), \bar{\phi}_{nlm_l}^{\alpha}(\zeta,\vec{k}), \bar{z}_{nlm_l}^{\alpha}(\zeta,\beta\theta\varphi). \quad (21)$$

For nonrelativistic scalar Slater orbitals

$$k_{nlm_l} = \chi_{nlm_l}(\zeta,\vec{r}), u_{nlm_l}(\zeta,\vec{k}), v_{nlm_l}(\zeta,\beta\theta\varphi). \quad (22)$$

The functions $\psi_{nlm_l}^{\alpha}, \bar{\psi}_{nlm_l}^{\alpha}, \phi_{nlm_l}^{\alpha}, \bar{\phi}_{nlm_l}^{\alpha}, z_{nlm_l}^{\alpha}, \bar{z}_{nlm_l}^{\alpha}, \chi_{nlm_l}, u_{nlm_l}$ and $v_{nlm_l}$ occurring in these formulas are determined by Eqs. (1)-(9).

The quasirelativistic and nonrelativistic functions satisfy the following orthonormality relations:

$$\int K_{njm_j}^{\alpha l^\dagger s}(\zeta,\vec{x})\bar{K}_{n'j'm_j'}^{\alpha l's}(\zeta,\vec{x})d\vec{x} = \delta_{nn'}\delta_{ll'}\delta_{jj'}\delta_{m_j m_j'} \quad (23)$$

$$\int K_{njm_j}^{l^\dagger s}(\zeta,\vec{x})K_{n'j'm_j'}^{l's}(\zeta,\vec{x})d\vec{x}$$
$$= \frac{(n+n')!}{\left[(2n)!(2n')!\right]^{1/2}}\delta_{ll'}\delta_{jj'}\delta_{m_j m_j'}, \quad (24)$$

$$\int k_{nlm_l(\lambda)}^{\alpha*}(\zeta,\vec{x})\bar{k}_{n'l'm_l'(\lambda)}^{\alpha}(\zeta,\vec{x})d\vec{x} = \delta_{nn'}\delta_{ll'}\delta_{m_l m_l'}, \quad (25)$$

$$\int k_{nlm_l(\lambda)}^{*}(\zeta,\vec{x})k_{n'l'm_l'(\lambda)}(\zeta,\vec{x})d\vec{x}$$
$$= \frac{(n+n')!}{\left[(2n)!(2n')!\right]^{1/2}}\delta_{ll'}\delta_{m_j m_j'}, \quad (26)$$

where $\vec{x} = \vec{r}, \vec{k}, \beta\theta\varphi$ and $d\vec{x} = d^3\vec{r}, d^3\vec{k}, d\Omega(\zeta,\beta\theta\varphi)$.

In order to construct the complete orthonormal sets of relativistic $2(2s+1)$-component tensor wave functions, and tensor Slater orbitals in coordinate, momentum and four-dimensional spaces we use the quasirelativistic $2s+1$-component tensor wave functions $\left(K_{njm_j}^{\alpha ls}, \bar{K}_{njm_j}^{\alpha ls} \text{ and } K_{njm_j}^{\alpha,l+t,s}, \bar{K}_{njm_j}^{\alpha,l+t,s}\right)$ and tensor Slater orbitals $\left(K_{njm_j}^{ls} \text{ and } K_{njm_j}^{l+t,s}\right)$ defined by Eqs. (14), (15) and (18). Here, the values of parameter t are determined from the relation

$$j = \begin{cases} l+t & \text{for } s=0,1,2,... \quad (-s \le t(1) \le s) \\ l+\frac{1}{2}t & \text{for } s=1/2,3/2,... \quad (-2s \le t(2) \le 2s) \end{cases} \quad (27)$$

Then, we finally establish the following formulas through the quasirelativistic and nonrelativistic functions:

For relativistic tensor wave functions

$$K_{njm_j}^{\alpha lst} = \frac{1}{\sqrt{2}}\begin{pmatrix} K_{njm_j}^{\alpha ls} \\ K_{njm_j}^{\alpha,l+t,s} \end{pmatrix} \quad (28a)$$

$$= \frac{1}{\sqrt{2}}\begin{pmatrix} a_{jm_j}^{ls}(0)k_{nlm_l(0)}^{\alpha} \\ a_{jm_j}^{ls}(1)k_{nlm_l(1)}^{\alpha} \\ \vdots \\ a_{jm_j}^{ls}(2s)k_{nlm_l(2s)}^{\alpha} \\ a_{jm_j}^{l+t,s}(0)k_{n,l+t,m_{l+t}(0)}^{\alpha} \\ a_{jm_j}^{l+t,s}(1)k_{n,l+t,m_{l+t}(1)}^{\alpha} \\ \vdots \\ a_{jm_j}^{l+t,s}(2s)k_{n,l+t,m_{l+t}(2s)}^{\alpha} \end{pmatrix} \quad (28b)$$

$$\bar{K}_{njm_j}^{\alpha lst} = \frac{1}{\sqrt{2}}\begin{pmatrix} \bar{K}_{njm_j}^{\alpha ls} \\ \bar{K}_{njm_j}^{\alpha,l+t,s} \end{pmatrix} \quad (29a)$$

$$= \frac{1}{\sqrt{2}}\begin{pmatrix} a_{jm_j}^{ls}(0)\bar{k}_{nlm_l(0)}^{\alpha} \\ a_{jm_j}^{ls}(1)\bar{k}_{nlm_l(1)}^{\alpha} \\ \vdots \\ a_{jm_j}^{ls}(2s)\bar{k}_{nlm_l(2s)}^{\alpha} \\ a_{jm_j}^{l+t,s}(0)\bar{k}_{n,l+t,m_{l+t}(0)}^{\alpha} \\ a_{jm_j}^{l+t,s}(1)\bar{k}_{n,l+t,m_{l+t}(1)}^{\alpha} \\ \vdots \\ a_{jm_j}^{l+t,s}(2s)\bar{k}_{n,l+t,m_{l+t}(2s)}^{\alpha} \end{pmatrix}. \quad (29b)$$

For relativistic tensor Slater orbitals

$$K_{njm_j}^{lst} = \frac{1}{\sqrt{2}}\begin{pmatrix} K_{njm_j}^{ls} \\ K_{njm_j}^{l+t,s} \end{pmatrix} \quad (30a)$$

$$= \frac{1}{\sqrt{2}}\begin{pmatrix} a_{jm_j}^{ls}(0)k_{nlm_l(0)} \\ a_{jm_j}^{ls}(1)k_{nlm_l(1)} \\ \vdots \\ a_{jm_j}^{ls}(2s)k_{nlm_l(2s)} \\ a_{jm_j}^{l+t,s}(0)k_{n,l+t,m_{l+t}(0)} \\ a_{jm_j}^{l+t,s}(1)k_{n,l+t,m_{l+t}(1)} \\ \vdots \\ a_{jm_j}^{l+t,s}(2s)k_{n,l+t,m_{l+t}(2s)} \end{pmatrix}. \quad (30b)$$

Thus, in coordinate, momentum and four-dimensional spaces we have

$2(2s+1)$ kinds of independent complete orthonormal sets of relativistic tensor wave functions, and relativistic tensor Slater orbitals. These functions satisfy the following orthogonality relations:

$$\int {}^{t}K_{njm_j}^{\alpha ls\dagger}(\zeta,\vec{x}){}^{t'}\overline{K}_{n'j'm_j'}^{\alpha l's}(\zeta,\vec{x})d\vec{x} = \delta_{nn'}\delta_{ll'}\delta_{jj'}\delta_{m_j m_j'}\delta_{tt'}$$
(31)

$$\int {}^{t}K_{njm_j}^{ls\dagger}(\zeta,\vec{x}){}^{t'}K_{n'j'm_j'}^{l's}(\zeta,\vec{x})d\vec{x}$$
$$= \frac{(n+n')!}{\left[(2n)!(2n')!\right]^{1/2}}\delta_{ll'}\delta_{jj'}\delta_{m_j m_j'}\delta_{tt'}, \quad (32)$$

where $\alpha = 1, 0, -1, -2, ...$

As can be seen from the formulas presented in this work, all of the relativistic tensor wave functions and tensor Slater orbitals are expressed through the corresponding nonrelativistic scalar functions defined in coordinate, momentum and four-dimensional spaces. Thus, the expansion and one-range addition theorems obtained in [18] for the $\psi^\alpha$-ETO, $\phi^\alpha$-MSO, $z^\alpha$-HSH and $\chi$-STO can be also used in the case of relativistic functions in coordinate, momentum and four-dimensional spaces.

## 3. Evaluation of Overlap Integrals over Relativistic Tensor Slater Orbitals in Coordinate Space

As an example of application, we evaluate in coordinate space the two-center overlap integrals over relativistic tensor Slater orbitals with the same screening parameters defined as

$${}^{tt'}S_{njm_j,n'j'm_j'}^{ls,l's}(\vec{G}) = \int {}^{t}X_{njm_j}^{ls\dagger}(\zeta,\vec{r}){}^{t'}X_{n'j'm_j'}^{l's}(\zeta,\vec{r}-\vec{R})d^3\vec{r}, \quad (33)$$

where $\vec{r} = \vec{r}_a, \vec{r}-\vec{R} = \vec{r}_b$, $\vec{R} = \vec{R}_{ab}$ and $\vec{G} = 2\zeta\vec{R}$. In order to evaluate these integrals we use Eq. (30b) for ${}^{t}K_{njm_j}^{ls} = {}^{t}X_{njm_j}^{ls}$. Then, we obtain for integral (33) the following relation in terms of nonrelativistic overlap integrals:

$${}^{tt'}S_{njm_j,n'j'm_j'}^{ls,l's}(\vec{G}) = \frac{1}{2}\sum_{\lambda=0}^{2s}\{a_{jm_j,j'm_j'}^{ls,l's}(\lambda)s_{nlm_l(\lambda),n'l'm_l'(\lambda)}(\vec{G}) + \quad (34)$$
$$a_{jm_j;j'm_j'}^{l+t,s;l'+t',s}(\lambda)s_{n,l+t,m_{l+t}(\lambda);n',l'+t',m_{l'+t'}'(\lambda)}(\vec{G})\}$$

where $a_{jm_j,j'm_j'}^{ls,l's}(\lambda) = a_{jm_j}^{ls}(\lambda)a_{j'm_j'}^{l's}(\lambda)$. The overlap integrals over scalar Slater orbitals occurring on the right-hand side of Eq. (34) are determined by

$$s_{nlm_l,n'l'm_l'}(\vec{G}) = \int \chi_{nlm_l}^*(\zeta,\vec{r})\chi_{n'l'm_l'}(\zeta,\vec{r}-\vec{R})d^3\vec{r} \quad (35a)$$

$$= \{[2(n+\alpha)]!/(2n)!\}^{1/2}$$
$$\sum_{\mu=l+1}^{n+\alpha}\sum_{\mu'=l'+1}^{n'}\frac{1}{(2\mu)^\alpha}\overline{\omega}_{n+\alpha,\mu}^{\alpha l}\overline{\omega}_{n'\mu'}^{\alpha l'}s_{\mu lm_l,\mu'l'm_l'}^\alpha(\vec{G}), \quad (35b)$$

where

$$s_{\mu lm_l,\mu'l'm_l'}^\alpha(\vec{G}) = \int \overline{\psi}_{\mu lm_l}^{\alpha*}(\zeta,\vec{r})\psi_{\mu'l'm'}^\alpha(\zeta,\vec{r}-\vec{R})d^3\vec{r}. \quad (36)$$

See Ref. [15] for the exact definition of coefficients $\overline{\omega}^{\alpha l}$. As we see from Eq.(35b), the overlap integral of scalar Slater orbitals are expressed in terms of overlap integrals over scalar $\psi^\alpha - ETO$. The analytical relations for the evaluation of nonrelativistic overlap integrals over $\psi^\alpha - ETO$ were obtained in [18].

The values of relativistic overlap integrals over tensor Slater orbitals with the same screening parameters obtained from the different complete sets of $\psi^\alpha - ETO$ $(\alpha = 1, 0, -1)$ using Mathematica 5.0 international mathematical software are presented in tables 1, 2 and 3. As we see from the tables that the suggested approach guarantees a highly accurate calculation of the overlap integrals over relativistic tensor Slater orbitals.

It should be noted that the overlap integrals over relativistic tensor Slater orbitals with the same screening parameters may be played a significant role in the calculation of arbitrary multicenter integrals arising in coordinate, momentum and four-dimensional spaces when Relativistic quantum mechanics is employed for the atomic, molecular and nuclear systems. Thus, the relations for the nonrelativistic overlap integrals over scalar $\psi^\alpha$-ETO, $\phi^\alpha$-MSO, $z^\alpha$-HSH and $\chi$-STO presented in our previous papers can be used in the evaluation of multicenter integrals over corresponding quasirelativistic and relativistic tensor wave functions and tensor Slater orbitals.

Table1. The values of overlap integrals over relativistic tensor Slater orbitals for $s=1/2$ obtained from the different complete sets of nonrelativistic $\psi^\alpha$-ETO in molecular coordinate system

| $n$ | $l$ | $t$ | $j$ | $m_j$ | $n'$ | $l'$ | $t'$ | $j'$ | $m'_j$ | $\theta$ | $\varphi$ | $G=2\zeta R$ | ${}^{tt'}S^{l\frac{1}{2},l'\frac{1}{2}}_{njm_j,n'j'm'_j}(\vec{G})$ | | |
|---|---|---|---|---|---|---|---|---|---|---|---|---|---|---|---|
| | | | | | | | | | | | | | $\alpha=1$ | $\alpha=0$ | $\alpha=-1$ |
| 3 | 1 | 1 | 3/2 | 1/2 | 2 | 1 | -1 | 1/2 | 1/2 | 0 | 0 | 30 | 5.3906333976E-05 | 5.3906333976E-05 | 5.3906333976E-05 |
| 5 | 3 | -1 | 5/2 | 3/2 | 4 | 1 | 1 | 3/2 | 3/2 | 0 | 0 | 25 | 1.7719515338E-03 | 1.7719515338E-03 | 1.7719515338E-03 |
| 6 | 3 | 1 | 7/2 | -3/2 | 3 | 2 | 1 | 5/2 | -3/2 | 0 | 0 | 50 | -1.6954454461E-05 | -1.6954454461E-05 | -1.6954454461E-05 |
| 7 | 5 | -1 | 9/2 | 5/2 | 6 | 4 | -1 | 7/2 | 5/2 | 0 | 0 | 60 | -9.5411661925E-06 | -9.5411661925E-06 | -9.5411661925E-06 |
| 9 | 6 | 1 | 13/2 | 7/2 | 8 | 5 | 1 | 11/2 | 7/2 | 0 | 0 | 80 | 3.9377128444E-07 | 3.9377128444E-07 | 3.9377128444E-07 |
| 2 | 0 | 1 | 1/2 | 1/2 | 2 | 0 | 1 | 1/2 | 1/2 | $\pi/4$ | $2\pi/3$ | 40 | 2.3403254297E-06 | 2.3403254297E-06 | 2.3403254297E-06 |
| 4 | 2 | -1 | 3/2 | -3/2 | 3 | 2 | 1 | 5/2 | 1/2 | $\pi/3$ | $4\pi/3$ | 55 | 5.8793612695E-08 | 5.8793612695E-08 | 5.8793612695E-08 |
| 5 | 3 | 1 | 7/2 | 5/2 | 4 | 2 | -1 | 3/2 | 1/2 | $\pi/6$ | $5\pi/4$ | 60 | 8.2690143328E-08 | 8.2690143328E-08 | 8.2690143328E-08 |
| 8 | 6 | -1 | 11/2 | -7/2 | 7 | 5 | -1 | 9/2 | 5/2 | $3\pi/5$ | $7\pi/6$ | 70 | -9.0245954747E-07 | -9.0245954747E-07 | -9.0245954747E-07 |
| 10 | 7 | 1 | 15/2 | 9/2 | 10 | 6 | 1 | 13/2 | 9/2 | $\pi/8$ | $\pi/4$ | 85 | 4.4074900820E-07 | 4.4074900820E-07 | 4.4074900820E-07 |

Table2. The values of overlap integrals over relativistic tensor Slater orbitals for $s=1$ obtained from the different complete sets of nonrelativistic $\psi^{\alpha}$-ETO in molecular coordinate system

| $n$ | $l$ | $t$ | $j$ | $m_j$ | $n'$ | $l'$ | $t'$ | $j'$ | $m'_j$ | $\theta$ | $\varphi$ | $G=2\zeta R$ | $^{tt'}S^{l1,l'1}_{njm_j,n'j'm'_j}(\vec{G})$ | | |
|---|---|---|---|---|---|---|---|---|---|---|---|---|---|---|---|
| | | | | | | | | | | | | | $\alpha=1$ | $\alpha=0$ | $\alpha=-1$ |
| 2 | 0 | 1 | 1 | 0 | 1 | 0 | 1 | 1 | 0 | 0 | 0 | 15 | 2.5775386500E-02 | 2.5775386500E-02 | 2.5775386500E-02 |
| 3 | 1 | 1 | 2 | 0 | 2 | 1 | -1 | 0 | 0 | 0 | 0 | 20 | 1.9523994419E-02 | 1.9523994419E-02 | 1.9523994419E-02 |
| 4 | 2 | 0 | 2 | 1 | 3 | 1 | 0 | 1 | 1 | 0 | 0 | 30 | -9.0790008925E-03 | -9.0790008925E-03 | -9.0790008925E-03 |
| 5 | 4 | -1 | 3 | 2 | 4 | 3 | 0 | 3 | 2 | 0 | 0 | 45 | 6.6927400519E-05 | 6.6927400519E-05 | 6.6927400519E-05 |
| 6 | 4 | 1 | 5 | 0 | 6 | 5 | -1 | 4 | 0 | 0 | 0 | 70 | 1.1418547340E-06 | 1.1418547340E-06 | 1.1418547340E-06 |
| 4 | 2 | 1 | 3 | -2 | 5 | 3 | 0 | 3 | 1 | $\pi/12$ | $\pi/6$ | 85 | -1.9428019254E-11 | -1.9428019254E-11 | -1.9428019254E-11 |
| 5 | 2 | 0 | 2 | 1 | 6 | 3 | 1 | 4 | -3 | $\pi/5$ | $2\pi/7$ | 25 | 1.5142546029E-02 | 1.5142546029E-02 | 1.5142546029E-02 |
| 7 | 6 | -1 | 5 | 4 | 7 | 5 | 1 | 6 | 5 | $3\pi/5$ | $5\pi/4$ | 40 | -6.0425312244E-06 | -6.0425312244E-06 | -6.0425312244E-06 |
| 9 | 8 | 0 | 8 | 6 | 4 | 3 | -1 | 2 | 1 | $4\pi/7$ | $\pi/3$ | 60 | 1.8078966695E-05 | 1.8078966695E-05 | 1.8078966695E-05 |
| 11 | 9 | 1 | 10 | 7 | 8 | 6 | 0 | 6 | 4 | $\pi/4$ | $\pi/4$ | 90 | -1.0143726361E-06 | -1.0143726361E-06 | -1.0143726361E-06 |

Table3. The values of overlap integrals over relativistic tensor Slater orbitals for $s=3/2$ obtained from the different complete sets of nonrelativistic $\psi^\alpha$-ETO in molecular coordinate system

| $n$ | $l$ | $t$ | $j$ | $m_j$ | $n'$ | $l'$ | $t'$ | $j'$ | $m'_j$ | $\theta$ | $\varphi$ | $G=2\zeta R$ | ${}^{tt'}S^{l\frac{3}{2},l'\frac{3}{2}}_{njm_j,n'j'm'_j}(\vec{G})$ | | |
|---|---|---|---|---|---|---|---|---|---|---|---|---|---|---|---|
| | | | | | | | | | | | | | $\alpha=1$ | $\alpha=0$ | $\alpha=-1$ |
| 3 | 1 | 3 | 5/2 | 1/2 | 2 | 1 | 1 | 3/2 | 1/2 | 0 | 0 | 35 | -1.7321448439E-03 | -1.7321448439E-03 | -1.7321448439E-03 |
| 4 | 3 | 1 | 7/2 | 3/2 | 3 | 1 | 3 | 5/2 | 3/2 | 0 | 0 | 50 | 1.4650668362E-05 | 1.4650668362E-05 | 1.4650668362E-05 |
| 5 | 2 | -3 | 1/2 | 1/2 | 5 | 1 | -1 | 1/2 | 1/2 | 0 | 0 | 60 | 2.9628984197E-06 | 2.9628984197E-06 | 2.9628984197E-06 |
| 7 | 5 | -1 | 9/2 | 5/2 | 6 | 3 | 1 | 7/2 | 5/2 | 0 | 0 | 80 | -1.2956401987E-10 | -1.2956401987E-10 | -1.2956401987E-10 |
| 9 | 6 | 3 | 15/7 | 11/2 | 9 | 8 | -3 | 13/2 | 11/2 | 0 | 0 | 90 | -2.8619877889E-10 | -2.8619877889E-10 | -2.8619877889E-10 |
| 3 | 1 | 3 | 5/2 | 1/2 | 2 | 1 | -1 | 1/2 | 1/2 | π/4 | π/4 | 40 | 5.4913839014E-06 | 5.4913839014E-06 | 5.4913839014E-06 |
| 5 | 3 | -1 | 5/2 | 3/2 | 4 | 1 | 1 | 3/2 | 3/2 | 3π/5 | 5π/4 | 55 | -4.9051537666E-06 | -4.9051537666E-06 | -4.9051537666E-06 |
| 8 | 6 | -1 | 11/2 | -7/2 | 7 | 5 | -1 | 9/2 | 5/2 | π/7 | π/3 | 70 | 3.5073321575E-07 | 3.5073321575E-07 | 3.5073321575E-07 |
| 9 | 5 | 3 | 13/2 | 9/2 | 8 | 4 | 3 | 11/2 | 11/2 | π/12 | 7π/6 | 85 | -6.1163164974E-10 | -6.1163164974E-10 | -6.1163164974E-10 |
| 10 | 7 | 1 | 15/2 | 13/2 | 10 | 6 | 1 | 13/2 | 9/2 | 2π/3 | 3π/4 | 100 | -5.8181215412E-09 | -5.8181215412E-09 | -5.8181215412E-09 |